\documentclass[prd,aps]{revtex4}
\usepackage{color}
\usepackage{amsfonts}
\usepackage{cancel}
\usepackage{mathrsfs}
\usepackage{cancel}
\usepackage{graphics}
\RequirePackage[reqno]{amsmath}
\begin{document}
\title{Abelian Ashtekar formulation from the ADM action}
\author{ Ernesto Contreras$^{1}$ and Lorenzo Leal$^{1}$
 $^{2}$}
\affiliation {1. Centro de F\'{\i}sica Te\'{o}rica y
Computacional, Facultad de Ciencias, Universidad Central de
Venezuela, AP 47270, Caracas 1041-A, Venezuela. \\
 2. Departamento
de F\'{\i}sica, Universidad Sim\'on Bol\'{\i}var,\\ Aptdo. 89000,
Caracas 1080-A, Venezuela.\\ }
\begin{abstract}
We study the Ashtekar formulation of linear gravity starting from the ADM first order action for the non linear theory, linearizing it, and  performing a canonical
transformation that coordinatizes  the phase space in terms of the already linearized Ashtekar variables. The results obtained in this way are in accordance with those
obtained through the standard method, in which, after introducing the Ashtekar variables for the full theory, a linearization around the flat Abelian connection and its
conjugate momentum is performed.
\end{abstract}
\maketitle

\section{Introduction.}
Linearized gravity is the starting point to study gravitational waves and weak gravitational field  phenomena\cite{misner,wald, carroll, deser,green,barnich_a,
 barnich_b,nieto,hennaux}.  Also, there are interesting formal relationships among electromagnetism and linear gravity, such as duality symmetry \cite{hennaux} 
  and gauge invariance \cite{nieto}, that constitute a permanent source of fruitful theoretical developments. On the other hand, the quantization of linearized gravity
results in quantum states that lie in the familiar graviton Fock space, on which the conventional perturbative approaches to quantum gravity
are based. Nevertheless, such approaches seem to fail due to the lack of renormalizability \cite{varadarajan}.\\

An alternative to build a quantum theory of gravity that avoids the renormalizability problems is provided by  Loop Quantum Gravity (LQG). This formulation may be obtained by  applying the Dirac method to  the Einstein theory in order to bring it into a canonical form, then, the phase space is parameterized by the Ashtekar variables
\cite{peldan,thiemann,rosas, baez, barbero_a,barbero_b,gaul,perez,ashtekar_a}. These variables are a $SU(2)$ connection $A^{i}_{a}$
and a densitised
triad $E^{a}_{i}$, which is a vector with respect to $SU(2)$ and a density of weight one \cite{barbero_a, barbero_b,gaul,perez} (here, the first letters of the alphabet
label space coordinates and those belonging to the middle of the alphabet are SU(2) indexes). The phase space becomes that of a conventional $SU(2)$ Yang-Mills theory except
by the fact that besides the
 Yang-Mills Gauss constraint, that generates $SU(2)$ gauge transformations, there appear two more constraints: the vectorial constraint generating $3$-dimensional space
diffeomorphisms and the scalar or Hamiltonian constraint, which generates time diffeomorphisms \cite{gaul,perez}. After casting gravity as a SU(2) Yang-Mills theory, the
introduction of non-canonical loop and area-dependent operators  allows to represent the constraints in a geometrical way that gives rise to the Loop Representation of
Quantum Gravity, or LQG \cite{baez, barbero_a,barbero_b,gaul,perez,ashtekar_a}.\\

Even though LQG is a promising candidate to achieve a quantum theory of gravity \cite{ashtekar_b,ashtekar_c, smolin,thieman_b}, some questions are still open as, for example,
how does flat spacetime arises from the full quantum theory. Moreover, understanding the relationship between the
quantum states of linearized gravity and those of the full non-perturbative loop quantum gravity, would shed light on the reasons behind
the failure of perturbative methods, as  pointed in reference \cite{varadarajan}. Hence, it is interesting to study the linearized theory in the Ahstekar variables.\\

The standard way to get a linearized theory of gravity in Ashtekar variables around flat space-time, is to start from the Ashtekar formulation, and to consider perturbations around the phase space
 point
$E^{a}_{i}=\delta^{a}_{i}$ and $A^{i}_{a}=0$  \cite{varadarajan, dibartolo,ashtekar_d}. After that, the phase space results to be
coordinatized by  three $U(1)$ connections (one for each value of $i$) $A^{i}_{a}$ and their  conjugate momenta $e^{a}_{i}$, which are  small perturbations
 around the ``flat'' triad $\delta^{a}_{i}$ \cite{varadarajan, dibartolo,ashtekar_d}. But there is another route that, as far as we know, has not been  discussed in detail
in the literature, and that we shall study in this paper. It consists in taking as starting point the linearized  $ADM$ action
 \cite{misner, peldan,thiemann,rosas,baez,arnowitt,giulini,franke} and performing a canonical transformation that  coordinatizes the phase space of the already
linear theory with three pairs of $U(1)$ connections and their conjugate momenta. As we shall see, the result obtained by this method coincides with that of the "standard" route,
\cite{varadarajan, dibartolo,ashtekar_d}. In short, one could say that the processes of "linearizing" and "converting into Ashtekar variables" commute.\\

The paper is organized as follow. In section $2$ we present an overview of the ADM formulation. In the next section we describe how to obtain
the linearized ADM action following standard methods. In section $4$ we perform a canonical transformation to parameterize
the phase space in terms of canonical pairs described in a non-coordinate basis. Then, in  section $5$ we introduce a linearized  spin connection and perform a second canonical transformation that finally yields a characterization of the phase space in terms of linearized Ashtekar variables. In the last section we present some concluding remarks.

\section{ADM formulation of gravity}
Eintein's equations can be derived through a variational principle from the Einstein-Hilbert action \cite{misner, wald,carroll,baez}
\begin{eqnarray}\label{uno}
S=\int\limits_{M}R\sqrt{-g}\, d^{4}x.
\end{eqnarray}
The space-time $M$ is an oriented semi-Riemannian manifold, $R$ is the Ricci scalar and $g$ is the determinant of the  metric tensor. A Hamiltonian formulation of general
relativity can be achived
 following the $ADM$ procedure \cite{misner, wald,baez} that we briefly summarize. Consider a separation of $M$ in a mono-parametric family of achronals hypersufaces
 $\Sigma_{\tau}$ with
$\tau\in\mathbb{R}$. Thereby, the dynamics may be described in terms of changes between successive hypersurfaces. More precisely, if $\gamma$ represents the world line of
$p\in\Sigma_{\tau}$ and $\partial_{\tau}$ is the vector field tangent to $\gamma$, the dynamics in $M$  is given by the changes in the $\partial_{\tau}$ direction suffered  by
quantities belonging to $\Sigma_{\tau}$ .
The vector field $\partial_{\tau}$ can be written down in terms of its tangent and normal vector fields components relative to $\Sigma_{\tau}$ \cite{wald,baez} as
\begin{eqnarray}\label{dos}
\partial_{\tau}=Nn+\vec{N},
\end{eqnarray}
where $N$ is  called the "lapse function"; $\vec{N}$, which belongs to the tangent space to $\Sigma_{\tau}$ at $p$, is the "shift" vector, and $n$ is
the unitary vector field normal to $\Sigma_{\tau}$ at $p$,
 obeying $g_{\mu\nu}n^{\mu}n^{\nu}=-1$ (in the case of Lorentzian $g_{\mu\nu}$) with $\mu,\nu=0,1,2,3$.\\

The metric components $g_{\mu\nu}$ can be written in terms of  $N$ and $\vec{N}$ and  the induced $3$-metric $q_{\mu\nu}$ on
$\Sigma_{\tau}$ as follows. We assign local coordinates $x^{0},x^{1},x^{2},x^{3}$ to $p\in\Sigma_{\tau}$ such that $x^{0}=\tau$. Then
$\partial_{0}=\partial_{\tau}$ and the vector fields $\partial_{1},\partial_{2},\partial_{3}$ are tangent to $\Sigma_{\tau}$ at $p$.
The components of the metric are then given by \cite{baez, arnowitt}

\begin{eqnarray}\label{tres}
g_{ab}&=&q_{ab}\nonumber\\
g^{ab}&=&q^{ab}-N^{-2}N^{a}N^{b}\nonumber\\
g_{0a}&=&N_{a}\nonumber\\
g^{0a}&=&N^{-2}N^{a}\nonumber\\
g_{00}&=&-(N^{2}-N_{a}N^{a})\nonumber\\
g^{00}&=&-N^{-2},
\end{eqnarray}
where $a,b=1,2,3$. From these expressions it can be seen that
\begin{eqnarray}\label{dtmi}
\sqrt{-g}=N\sqrt{q},
\end{eqnarray}
where $q$ is the determinant of the induced $3$-metric $q_{ab}$ \cite{arnowitt,wald,baez}.\\

The Ricci scalar can be written as \cite{misner,wald,baez}
\begin{eqnarray}\label{cuatro}
R=\ ^{3}R-K^{ab}K_{ab}+K^{2},
\end{eqnarray}
where $^{3}R$ is the  Ricci scalar induced in $\Sigma_{\tau}$, defined by
\begin{eqnarray}\label{siete}
^{3}R=\delta_{ab}q^{cd}\ ^{3}R^{a}_{cbd},
\end{eqnarray}
with $^{3}R^{a}_{cbd}$ being the induced curvature, given in terms of the induced affine connection $^{3}\Gamma^{a}_{bc}$  by
\begin{eqnarray}\label{ocho}
^{3}R^{a}_{bcd}=\partial_{b}\ ^{3}\Gamma^{a}_{cd}-\partial_{c}\ ^{3}\Gamma^{a}_{bd}+\ ^{3}\Gamma^{e}_{cd}\ ^{3}\Gamma^{a}_{be}-\ ^{3}\Gamma^{e}_{bd}\ ^{3}\Gamma^{a}_{ce}.
\end{eqnarray}
In turn, the induced affine connection $^{3}\Gamma^{a}_{bc}$ can be written as
\begin{eqnarray}\label{nueve}
^{3}\Gamma^{a}_{bc}=\frac{1}{2}q^{ad}(\partial_{b}q_{cd}+\partial_{c}q_{bd}-\partial_{d}q_{bc}).
\end{eqnarray}

In equation \eqref{cuatro} also appears the $(0,2)$ extrinsic curvature tensor $K_{ab}$  \cite{misner,wald,baez}
\begin{eqnarray}\label{cinco}
K_{ab}=\frac{1}{2}N^{-1}(\partial_{0}q_{ab}-^{3}\nabla_{a}N_{b}-^{3}\nabla_{b}N_{a}),
\end{eqnarray}
and its trace $K=K_{ab}q^{ab}$. Additionally  $^{3}\nabla$ is the covariant derivative operator preserving the $3$-metric  \cite{wald},
which can be written down in terms of the "full" covariant derivative $\nabla$  as
\begin{eqnarray}\label{seis}
^{3}\nabla_{\mu}A_{\nu}=q^{\rho}_{\mu}q^{\sigma}_{\nu}\nabla_{\rho}A_{\sigma},
\end{eqnarray}
for every covariant vector field $A_{\mu}$  in $M$. \\

Replacing  (\ref{dtmi}) and (\ref{cuatro}) in (\ref{uno}) we can rewrite the action as
\begin{eqnarray}\label{diez}
S=\int\limits_{M}d^{4}x\mathcal{L}=\int\limits_{M}d^{4}x\sqrt{q}N(^{3}R-K^{ab}K_{ab}+K^{2}).
\end{eqnarray}
It should be underlined that this expression depends exclusively on quantities relative to the "space" $\Sigma_{\tau}$, and on the lapse and shift fields.
This fact facilitates the passage to the Hamiltonian formulation. Applying the Dirac canonical procedure to the action (\ref{diez})
we obtain the first order $ADM$ action \cite{misner,wald,baez,arnowitt}
\begin{eqnarray}\label{accionpo}
S=\int d^{4}x \, p^{ab}\dot{q_{ab}}-\int dt \, H &=&\int d^{4}x[p^{ab}\dot{q}_{ab}-
N\sqrt{q}(-^{3}R+K^{a}_{a}K^{b}_{b}-K^{ab}K_{ab})-N_{b}(-2 \ ^{3}\nabla_{a}p^{ab})],
\end{eqnarray}
where
\begin{eqnarray}\label{once}
p^{ab}=\frac{\partial\mathcal{L}}{\partial \dot{q}_{ab}}=-\sqrt{q}(K^{ab}-q^{ab}K)
\end{eqnarray}
is the  momentum conjugate to $q_{ab}$ and
\begin{eqnarray}\label{doce}
H=\int d^{3}x[N\mathcal{S}+N_{b}V^{b} ]
\end{eqnarray}
is the Hamiltonian, with
\begin{eqnarray}\label{trece}
\mathcal{S}&=&\sqrt{q}(-^{3}R+K^{ab}K_{ab}-K^{2})\approx0\nonumber\\
V^{b}&=&-2\ ^{3}\nabla_{a}p^{ab}\approx0,
\end{eqnarray}
being the scalar and vectorial constraints respectively \cite{misner,wald,baez,arnowitt}.
The lapse and the shift are then Lagrange multipliers enforcing the constraints.
These constraints are first class in Dirac's sense, and generate time and spatial diffeomorphisms, respectively, on the phase space.
The true dynamical variables are the spatial metric and its canonical conjugate, whose equations of motion are  given by
\begin{eqnarray}\label{catorce}
\dot{q}_{ab}&=&\{q_{ab},H\},\nonumber\\
\dot{p}^{ab}&=&\{p^{ab},H\}.
\end{eqnarray}
The fundamental Poisson brackets are
\begin{eqnarray}\label{dieciseis}
\{q_{ab}(x),p^{a'b'}(y)\}&=&(\delta^{a'}_{a}\delta^{b'}_{b}+\delta^{a'}_{b}\delta^{b'}_{a})\delta^{3}(x-y)\nonumber\\
\{p^{ab}(x),p^{a'b'}(y)\}&=&\{q_{ab}(x),q_{a'b'}(y)\}=0.
\end{eqnarray}
The canonical equations \eqref{catorce}, together with the scalar and vector constraints reproduce Einstein equations, as can be verified.

\section{ADM linearized gravity}
In order to linearize the theory, we  consider small perturbations $h_{ab}$ around the flat metric $\eta_{ab}$
\begin{eqnarray}\label{dieciochou}
q_{ab}&=&\eta_{ab}+h_{ab}\nonumber\\
q^{ab}&=&\eta^{ab}-h^{ab};\hspace*{30pt}h_{ab}<<1,
\end{eqnarray}
so that up to first order in $h_{ab}$ we have \cite{misner, carroll}
\begin{eqnarray}
q_{ab}q^{ac}=\delta^{c}_{b}.
\end{eqnarray}
The induced $3$-metric determinant, up to first order in the perturbation $h_{ab}$, is then given by
\begin{eqnarray}\label{dete}
q&=&=1+h,
\end{eqnarray}
where $h=\eta^{ab}h_{ab}$ is the trace of $h_{ab}$. In the linearized theory, indexes are raised and lowered with $\eta_{ab}$ and $\eta^{ab}$ rather than
$g_{ab}$ and $g^{ab}$. The linearized lapsus and shift are given by
\begin{eqnarray}\label{ventiuno}
N&=&1+\nu;\hspace*{20pt}\nu<<1\nonumber\\
N_{a}&=&\nu_{a}; \hspace*{30pt}\nu_{a}<<1,
\end{eqnarray}
as can be seen from their relationship with the metric components (equations \eqref{tres}).
Replacing (\ref{dieciochou}) and (\ref{ventiuno}) in (\ref{cinco}), recalling the "exact" expression for the covariant derivative of a $1-form$
\begin{eqnarray}
^{3}\nabla_{a}\nu_{b}=\partial_{a}\nu_{b}-\ ^{3}\Gamma^{c}_{ab}\nu_{c},
\end{eqnarray}
and substituting the linearized Christoffel symbols  \cite{carroll}
\begin{eqnarray}\label{simb}
^{3}\Gamma^{a}_{b c}&=&\frac{1}{2}(\partial_{b}h^{a}\ _{c}+\partial_{k}h^{a}\ _{b}-\partial^{a}h_{bc}),
\end{eqnarray}
we obtain the extrinsic curvature
\begin{eqnarray}\label{extrinseca}
K_{ab}=\frac{1}{2}(\partial_{0}h_{ab}-\partial_{a}\nu_{b}-\partial_{b}\nu_{a}),
\end{eqnarray}
up to first order in $h_{ab}$. Also, substituting (\ref{extrinseca}) in (\ref{once}) we obtain the linearized conjugate momenta
(\ref{once})
\begin{eqnarray}\label{mcl}
p_{ab}=-(K_{ab}-\eta_{ab}K)=-\frac{1}{2}(\partial_{0}h_{ab}-\eta_{ab}\partial_{0}h)+\frac{1}{2}(\partial_{a}\nu_{b}+\partial_{b}\nu_{a}-2\eta_{ab}\partial^{c}\nu_{c}).
\end{eqnarray}

To obtain the linearized equations of motion we have to keep terms up to second order in $h_{ab}$ in the the $ADM$ action.
The second order  Ricci scalar can be found from  equation (\ref{siete})
\begin{eqnarray}\label{escalarricci}
^{3}R=\delta_{ac}q^{bd}\ ^{3}R^{a}_{bcd}=\delta_{ac}(\eta^{bd}-h^{bd})\ ^{3}R^{a}_{bcd},
\end{eqnarray}
where in the exact expression for the Riemman tensor
\begin{eqnarray}
^{3}R^{a}_{bcd}=\partial_{b}\ ^{3}\Gamma^{a}_{cd}-\partial_{c}\ ^{3}\Gamma^{a}_{bd}+^{3}\Gamma^{e}_{cd}\ ^{3}\Gamma^{a}_{be}-^{3}\Gamma^{e}_{bd}\ ^{3}\Gamma^{a}_{ce},
\end{eqnarray}

we have to substitute (\ref{simb}). After that substitution we obtain
\begin{eqnarray}
^{3}R^{a}_{bcd}&=&\frac{1}{2}(\partial_{b}\partial_{d}h^{a}_{c}-\partial^{a}\partial_{b}h_{cd})
-\frac{1}{2}(\partial_{c}\partial_{d}h^{a}_{b}-\partial_{c}\partial^{a}h_{bd})
+\frac{1}{4}(\partial_{c}h^{e}_{d}+\partial_{d}h^{e}_{c}-\partial^{e}h_{cd})
(\partial_{b}h^{a}_{e}+\partial_{e}h^{a}_{b}-\partial^{a}h_{be})\nonumber\\
&-&\frac{1}{4}(\partial_{b}h^{e}_{d}+\partial_{d}h^{e}_{b}-\partial^{e}h_{bd})
(\partial_{c}h^{a}_{e}+\partial_{e}h^{a}_{c}-\partial^{a}h_{ce}),
\end{eqnarray}
where we have kept  terms up to second order in the perturbation $h_{bd}$. Replacing the above equation in (\ref{escalarricci}) we obtain, to the desired order
\begin{eqnarray}
^{3}R&=&q^{bd}\ ^{3}R^{a}_{bad}=(\eta^{bd}-h^{bd})\ ^{3}R^{a}_{bad}=\partial^{a}\partial_{a}h-\partial^{a}\partial^{b}h_{ab}
-\frac{1}{2}\partial_{a}h^{b}_{c}\partial^{c}h^{a}_{b}
+\frac{1}{4}\partial^{a}h_{bc}\partial_{a}h^{bc}
+\frac{1}{4}\partial^{a}h\partial_{a}h.
\end{eqnarray}
Here we have already neglected those terms that will produce either cubic contributions or total derivative contributions to the action. After this we are ready
to write down the term $N\sqrt{^{3}q}\ ^{3}R$ up to second order in the perturbation:
\begin{eqnarray}\label{ngr}
N\sqrt{^{3}q}\ ^{3}R&=&(1+\nu)(1+\frac{1}{2}h)(\partial^{a}\partial_{a}h-\partial^{a}\partial^{b}h_{ab}
-\frac{1}{2}\partial_{a}h^{b}_{c}\partial^{c}h^{a}_{b}
+\frac{1}{4}\partial^{a}h_{bc}\partial_{a}h^{bc}
+\frac{1}{4}\partial^{a}h\partial_{a}h)\nonumber\\
&=&
-\frac{1}{2}\partial_{a}h^{b}_{c}\partial^{c}h^{a}_{b}
+\frac{1}{4}\partial^{a}h_{bc}\partial_{a}h^{bc}
-\frac{1}{4}\partial^{a}h\partial_{a}h+\frac{1}{2}\partial^{a}h\partial^{b}h_{ab}
+\nu(\partial^{a}\partial_{a}h-\partial^{a}\partial^{b}h_{ab})\nonumber\\
&=&\mathfrak{T}+\nu(\partial^{a}\partial_{a}h-\partial^{a}\partial^{b}h_{ab}),
\end{eqnarray}
where we have defined
\begin{eqnarray}
\mathfrak{T}= -\frac{1}{2}\partial_{a}h^{b}_{c}\partial^{c}h^{a}_{b}
+\frac{1}{4}\partial^{a}h_{bc}\partial_{a}h^{bc}
-\frac{1}{4}\partial^{a}h\partial_{a}h+\frac{1}{2}\partial^{a}h\partial^{b}h_{ab},
\end{eqnarray}
and neglected total derivative terms.
Finally, the contribution coming from the term $K_{ab}K^{ab}-K^{2}$ can be written as a function of the linearized  conjugate momentum as
\begin{eqnarray}\label{kmk}
K_{ab}K^{ab}-K^{2}=p^{ab}p_{ab}-\frac{1}{2}p^{2}.
\end{eqnarray}

Substituting (\ref{dieciochou}), (\ref{dete}), (\ref{ventiuno}), (\ref{ngr}) and (\ref{kmk}) in (\ref{accionpo}) we obtain the cuadratic action
(that provides the linearized canonical equations of motion) as
\begin{eqnarray}\label{accionpol}
S&=&\int d^{4}x(p^{ab}\dot{q}_{ab}
+p^{ab}p_{ab}-\frac{1}{2}p^{2}+\mathfrak{T}-\nu(\partial^{a}\partial^{b}h_{ab}-\partial^{a}\partial_{a}h)+2\nu_{a}\partial_{b}p^{ba}).
\end{eqnarray}
From this expression we read the linearized scalar and vectorial  constraints, which  are
\begin{eqnarray}\label{lgl}
\mathcal{S}&=&\partial^{a}\partial^{b}h_{ab}-\partial^{a}\partial_{a}h\nonumber\\
V^{b}&=&-2\partial_{a}p^{ab}.
\end{eqnarray}
Defining
\begin{eqnarray}
\mathcal{H}=-(p^{ab}p_{ab}-\frac{1}{2}p^{2}+\mathfrak{T}),
\end{eqnarray}
the linearized  action can be written down as
\begin{eqnarray}\label{apol}
S=\int d^{4}x(p^{ab}\dot{h}_{ab}-\mathcal{H}-\nu\mathcal{S}-\nu_{a}V^{a}).
\end{eqnarray}
From the above equation we see that the Hamiltonian of the theory is
\begin{eqnarray}
H=\int d^{3}x(\mathcal{H}+\nu\mathcal{S}+\nu_{a}V^{a}).
\end{eqnarray}
Thereby, the dynamics of the linearized theory is given by
\begin{eqnarray}
\dot{p}^{ab}&=&\{p^{ab},H\}\nonumber\\
\dot{q}_{ab}&=&\{h_{ab},H\},
\end{eqnarray}
with
\begin{eqnarray}\label{alg}
\{q_{ab}(x),p^{a'b'}(y)\}&=&(\delta^{a'}_{a}\delta^{b'}_{b}+\delta^{a'}_{b}\delta^{b'}_{a})\delta^{3}(x-y)\nonumber\\
\{p^{ab}(x),p^{a'b'}(y)\}&=&\{q_{ab}(x),q_{a'b'}(y)\}=0,
\end{eqnarray}
being the canonical algebra obeyed by the linearized variables, which is readily obtained from  equations (\ref{dieciseis}).

\section{Non Coordinate Basis}
Until now we have been working in a coordinate basis $\partial_{a}$ ($a=1,2,3$) of the tangent space at $p\in\Sigma_{\tau}$;  however, we can also associate to each $p$
a non coordinate basis  ${e}_{i}$  ($i=1,2,3$) \cite{carroll,baez}. Non coordinate basis play an important role in the Ashtekar formulation of gravity, whose linearized version is our objective.
These two basis are related by
\begin{eqnarray}
\mathfrak{e}^{i}_{a}e_{i}=\partial_{a},
\end{eqnarray}
Following the usual practice, we shall refer both to the basis vectors $e_{i}$ and to the components  $\mathfrak{e}^{i}_{a}$  of the coordinate basis
in the new one as the triad ($\mathfrak{e}^{a}_{i}$  is then the inverse triad: $\mathfrak{e}^{i}_{b}\mathfrak{e}^{a}_{i}=\delta^{a}_{b}$,
$\mathfrak{e}^{a}_{j}\mathfrak{e}^{i}_{a}=\delta^{i}_{j}$). The scalar product of vectors
is given by
\begin{eqnarray}
q(\partial_{a},\partial_{b})=q_{ab}=q(\mathfrak{e}^{i}_{a}e_{i},\mathfrak{e}^{j}_{b}e_{j})=\mathfrak{e}^{i}_{a}\mathfrak{e}^{j}_{b}q(e_{i},e_{j}),
\end{eqnarray}
hence, if the non coordinate basis is orthonormal  ($q(e_{i},e_{j})=\eta_{ij}$, $\eta_{ij}$ being the Euclidean metric) we shall have
\begin{eqnarray}\label{diecinueve}
q_{ab}=\mathfrak{e}^{i}_{a}\mathfrak{e}^{j}_{b}\eta_{ij}
\end{eqnarray}
and
\begin{eqnarray}\label{veinte}
\eta_{ij}=\mathfrak{e}^{a}_{i}\mathfrak{e}^{b}_{j}q_{ab}.
\end{eqnarray}

The densitized triad $E^{a}_{i}$ is defined as
\begin{eqnarray}\label{ventidos}
E^{a}_{i}=\mathfrak{e}\mathfrak{e}^{a}_{i},
\end{eqnarray}
with $(det\mathfrak{e}^{i}_{a})^{2}=\mathfrak{e}^{2}=q$. From this we have
\begin{eqnarray} \label{qqee}
qq^{ab}=E^{a}_{i}E^{b}_{j}\eta^{ij}.
\end{eqnarray}
In order to achieve our goal of obtaining linearized gravity in Ahstekar variables, we have first to write (\ref{diecinueve}), (\ref{veinte})
and (\ref{qqee}) in terms of the  linearized metric to get the linearized densitized triad. To this end we observe that from its definition,
the densitized triad should be written as
\begin{eqnarray}\label{tdl}
E^{a}_{i}=\delta^{a}_{i}+e^{a}_{i};
\end{eqnarray}
then, substituting  in (\ref{qqee})  and keeping terms up to second order in the perturbation  $e^{a}_{i}$ of the densitized triad we obtain
\begin{eqnarray*}
qq^{ab}=E^{a}_{i}E^{b}_{j}\eta^{ij}\Rightarrow(1+h)(\eta^{ab}-h^{ab})=(\delta^{a}_{i}+e^{a}_{i})(\delta^{b}_{j}+e^{b}_{j})\eta^{ij}
=\delta^{a}_{i}\delta^{b}_{j}\eta^{ij}+\delta^{a}_{i}e^{b}_{j}\eta^{ij}+\delta^{b}_{j}e^{a}_{i}\eta^{ij}.
\end{eqnarray*}
Hence
\begin{eqnarray}\label{ehe}
-h^{ab}+h\eta^{ab}&=&\delta^{a}_{i}e^{b}_{j}\eta^{ij}+\delta^{b}_{j}e^{a}_{i}\eta^{ij}.
\end{eqnarray}

Taking the time derivative of (\ref{ehe}) we get
\begin{eqnarray}
p_{ab}\dot{h}^{ab}&=&-(K_{ab}-\eta_{ab}K)\dot{h}^{ab}=-(K_{ab}\dot{h}^{ab}-\eta_{a'b'}K_{ab}\eta^{ab}\dot{h}^{a'b'})=
-K_{ab}(\dot{h}^{ab}-\eta^{ab}\dot{h})=K_{ab}(\delta^{a}_{i}\dot{e}^{b}_{j}+\delta^{b}_{j}\dot{e}^{a}_{i})\eta^{ij}\nonumber\\
&=&2K_{ab}\delta^{a}_{i}e^{b}_{j}\eta^{ij}=e^{b}_{j}\dot{k^{j}_{b}},
\end{eqnarray}
where we have defined
\begin{eqnarray}\label{ce}
k^{j}_{b}=-2K_{ab}\delta^{a}_{i}\eta^{ij},
\end{eqnarray}
and a total time derivative has been neglected because  this expression is going to be substituted in the action \eqref{apol}, and
total derivatives in the Lagrange density do not affect the equations of motion.

It can be shown that
\begin{eqnarray}\label{algebraek}
\{e^{a}_{i}(x),k^{j}_{b}(y)\}&=&\delta^{a}_{b}\delta^{j}_{i}\delta^{3}(x-y)\nonumber\\
\{e^{a}_{i}(x),e^{b}_{j}(y)\}&=&\{k^{i}_{a}(x),k^{j}_{b}(y)\}=0,
\end{eqnarray}
hence,  $e^{a}_{i}$ and $k^{i}_{a}$  form a new set of canonical variables. In what follows, we will calculate all the quantities needed to
write the first order action (\ref{apol}) in terms of these new canonical variables.  From (\ref{ehe}), we have
\begin{eqnarray}\label{th}
h=\delta^{i}_{a}e^{a}_{i}=\delta^{a}_{i}e^{i}_{a}.
\end{eqnarray}
Replacing this in (\ref{ehe}) we obtain
\begin{eqnarray}\label{hlia}
h^{ab}=-\delta^{a}_{i}e^{b}_{j}\eta^{ij}-\delta^{b}_{j}e^{a}_{i}\eta^{ij}+\delta^{i}_{c}e^{c}_{i}\eta^{ab}.
\end{eqnarray}
Now, from equations (\ref{ventidos}) and  (\ref{tdl}) we obtain the linearized triad $\mathfrak{e}^{a}_{i}$
\begin{eqnarray}\label{tnil}
\mathfrak{e}^{a}_{i}
&=& \delta^{a}_{i}+e^{a}_{i}-\frac{1}{2}e^{b}_{j}\delta^{j}_{b}\delta^{a}_{i},
\end{eqnarray}
so that its inverse $\mathfrak{e}^{i}_{a}$ is given by
\begin{eqnarray}\label{til}
\mathfrak{e}^{i}_{a}
&=&\delta^{i}_{a}
-\delta^{j}_{a}\delta^{i}_{b}e^{b}_{j}
+\frac{1}{2}\delta^{i}_{a}\delta^{j}_{b}e^{b}_{j}.
\end{eqnarray}

Replacing (\ref{til}) in (\ref{veinte}) we can write the linearized metric $q_{ab}$ as
\begin{eqnarray}
q_{ab}&=&\mathfrak{e}^{i}_{a}\mathfrak{e}^{j}_{b}\eta_{ij}=\eta_{ab}
-\eta_{ac}\delta^{k}_{b}e^{c}_{k}
-\eta_{bc}\delta^{k}_{a}e^{c}_{k}
+\eta_{ab}\delta^{k}_{c}e^{c}_{k},
\end{eqnarray}
up to first order in $e^{a}_{i}$. Comparing this expression with $q_{ab}=\eta_{ab}+h_{ab}$, we read that
\begin{eqnarray}\label{hia}
h_{ab}=-\eta_{ac}\delta^{k}_{b}e^{c}_{k}
-\eta_{bc}\delta^{k}_{a}e^{c}_{k}
+\eta_{ab}\delta^{k}_{c}e^{c}_{k},
\end{eqnarray}
hence, substituting (\ref{hia}) and (\ref{th}) in (\ref{lgl}) we obtain the scalar constraint
\begin{eqnarray}\label{le}
\mathcal{S}=-2\partial^{b}\delta^{k}_{b}\partial_{c}e^{c}_{k}\approx0.
\end{eqnarray}
In turn, replacing (\ref{ce}) in (\ref{lgl}) we also obtain the vectorial constraint
\begin{eqnarray}\label{lv}
V_{a}&=&\delta^{c}_{i}\partial_{c}k^{i}_{a}-\delta^{b}_{i}\partial_{a}k^{i}_{b}\approx0.
\end{eqnarray}

Since the extrinsic curvature is a symmetric tensor, and this is not reflected in the Poisson algebra  (\ref{algebraek}), we must force this symmetry
in the form of another constraint
\begin{eqnarray}
\varepsilon^{abc}K_{ab}=0,
\end{eqnarray}
so that, substituting equation (\ref{ce}) in the above expression we obtain
\begin{eqnarray}
\varepsilon^{ikl}\delta^{a}_{k}\eta_{ji}k^{j}_{a}=0,\label{lgauss}
\end{eqnarray}
which is going to be the linearized Gauss constraint $G^{l}$ when we arrive to the Ashtekar variables. This new constraint, together with equations (\ref{le})
and (\ref{lv}), forms a set of first class constraints.\\

Using equations
(\ref{kmk}) and (\ref{ce}) we can rewrite the term $p^{ab}p_{ab}-\frac{1}{2}p^{2}$ as
\begin{eqnarray}\label{pth}
p^{ab}p_{ab}-\frac{1}{2}p^{2}=-\frac{1}{4}\delta^{a}_{i}\delta^{b}_{j}(k^{i}_{a}k^{j}_{b}-k^{i}_{b}k^{j}_{a}).
\end{eqnarray}

Putting all this together in the action  (\ref{accionpol}) we obtain
\begin{eqnarray}\label{aponv}
S=\int d^{4}x(e^{a}_{i}\dot{k}^{i}_{a}+\frac{1}{4}\delta^{a}_{j}\delta^{a'}_{j'}(k^{j}_{a'}k^{j'}_{a}
-k^{j}_{a}k^{j'}_{a'})+\mathfrak{T}
-\nu\mathcal{S}-\nu_{a}V^{a}-N_{i}G^{i})
\end{eqnarray}
(the explicit expression for  $\mathfrak{T}$ as a function of $e^{a}_{i}$  is unnecessary at this point, and we omit it for the sake of brevity).
Moreover
\begin{eqnarray}
\mathcal{S}&=&-2\partial^{b}\delta^{k}_{b}\partial_{c}e^{c}_{k}\approx0\nonumber\\
V_{a}&=&\delta^{c}_{i}\partial_{c}k^{i}_{a}-\delta^{b}_{i}\partial_{a}k^{i}_{b}\approx0\nonumber\\
G^{l}&=&\varepsilon^{ikl}\delta^{a}_{k}\eta_{ji}k^{j}_{a}\approx0.
\end{eqnarray}

From the first order action we read, besides the constraints, the Hamiltonian of the theory
\begin{eqnarray}
H=-\int d^{3}x(-\frac{1}{4}\delta^{a}_{j}\delta^{a'}_{j'}(k^{j}_{a'}k^{j'}_{a}
-k^{j}_{a}k^{j'}_{a'})-\mathfrak{T}
+\nu\mathcal{S}+\nu_{a}V^{a}+N_{i}G^{i}).
\end{eqnarray}
The equations of motion are given by
\begin{eqnarray}
\dot{e}^{a}_{i}&=&\{e^{a}_{i},H\}\nonumber\\
\dot{k}^{i}_{a}&=&\{k^{i}_{a},H\},
\end{eqnarray}
with
\begin{eqnarray}
\{e^{a}_{i}(x),k^{j}_{b}(y)\}&=&\delta^{a}_{b}\delta^{j}_{i}\delta^{3}(x-y)\nonumber\\
\{e^{a}_{i}(x),e^{b}_{j}(y)\}&=&\{k^{i}_{a}(x),k^{j}_{b}(y)\}=0
\end{eqnarray}
being the canonical algebra. The dynamics must be complemented with the constraints \eqref{le}, \eqref{lv} and \eqref{lgauss}. We have then attained a description
of linear gravity  in terms of the linearized densitized triad and the linearized extrinsic curvature as conjugate variables, starting from the ADM formulation of
linearized gravity. This setting provides a starting point to obtain the Ashtekar formulation of linearized gravity, which is our next step.

\section{Ashtekar Variables}
In order to carry out the canonical transformation towards the linearized Ahstekar variables we must introduce the linearized spin connection,
which allows to express the covariant derivative in the non coordinate basis. To this end let us recall some definitions of the "full" non-linear theory.
If  $\mathcal{A}=\mathcal{A}^{i}e_{i}$ is a contravariant vector field, we have \cite{carroll}
\begin{eqnarray}
\nabla \mathcal{A}&=&(\nabla_{a}\mathcal{A}^{i})dx^{a}\otimes e_{i}=(\partial_{a}\mathcal{A}^{i}+\varepsilon_{j}\ ^{i}\ _{k}\Gamma^{k}_{a}
\mathcal{A}^{j})dx^{a}\otimes e_{i},
\end{eqnarray}
where $\Gamma^{k}_{a}$ is the spin connection, that can be thought as a $SU(2)$ connection \cite{wald,baez,gaul,perez}.
To relate the spin connection with the affine connection we rewrite the last expression as
\begin{eqnarray}\label{dcce}
\nabla \mathcal{A}&=&(\nabla_{a}\mathcal{A}^{i})dx^{a}\otimes e_{i}=(\partial_{a}\mathcal{A}^{i}+\varepsilon_{j}\ ^{i}\ _{k}\Gamma^{k}_{a}
\mathcal{A}^{j})dx^{a}\otimes e_{i}\nonumber\\
&=&(\partial_{a}(\mathfrak{e}^{i}_{b}\mathcal{A}^{b})+\varepsilon_{j}\ ^{i}\ _{k}\Gamma^{k}_{a}\mathfrak{e}^{j}_{b}\mathcal{A}^{b})dx^{a}
\otimes(\mathfrak{e}^{c}_{i}\partial_{c})
\nonumber\\
&=&(\partial_{a}\mathcal{A}^{c}+\mathcal{A}^{b}(\mathfrak{e}^{c}_{i}\partial_{a}\mathfrak{e}^{i}_{b}
+\mathfrak{e}^{c}_{i}\varepsilon_{j}\ ^{i}\ _{k}\Gamma^{k}_{a}\mathfrak{e}^{j}_{b}))dx^{a}\otimes\partial_{c}.
\end{eqnarray}

On the other hand, in the coordinate basis one has
\begin{eqnarray}\label{dcca}
\nabla \mathcal{A}=(\nabla_{a}\mathcal{A}^{b})dx^{a}\otimes\partial_{b}=(\partial_{a}\mathcal{A}^{b}+\ ^{3}\Gamma^{b}_{ac}\mathcal{A}^{c})dx^{a}\otimes\partial_{b},
\end{eqnarray}
where $^{3}\Gamma^{a}_{bc}$ are the Christoffel symbols \cite{wald,carroll}. Comparing both expressions
 we obtain
\begin{eqnarray}\label{dcce}
\mathfrak{e}^{c}_{i}\partial_{a}\mathfrak{e}^{i}_{b}+\varepsilon_{j}\ ^{i}\ _{k}\mathfrak{e}^{c}_{i}\mathfrak{e}^{j}_{b}\Gamma^{k}_{a}=\ ^{3}\Gamma^{c}_{ab}.
\end{eqnarray}

The linearized version of this expression is obtained with the aid of equations (\ref{tnil}) and (\ref{til})
\begin{eqnarray}\label{celi}
\varepsilon_{j}\ ^{i}\ _{k}\Gamma^{k}_{a}&=&\delta_{c}^{i}\delta^ {b}_{j}\ ^{3}\Gamma^{c}_{ab}+
\delta_{c}^{i}\partial_{a}e^{c}_{j}-\frac{1}{2}\delta_{j}^{i}\delta^{k}_{d}\partial_{a}e^{d}_{k},
\end{eqnarray}
where $\ ^{3}\Gamma^{c}_{ab}$ must be substituted by the linearized Christoffel symbols given by equation (\ref{simb}), while  $\Gamma^{i}_{a}$
is now the linearized spin connection. From this expression we have
\begin{eqnarray}\label{lgav}
\varepsilon_{i}\ ^{j}\ _{k}\Gamma^{k}_{a}\delta^{a}_{j}&=&
(\delta^{j}_{b}\delta^{c}_{i}\ ^{3}\Gamma^{b}_{ac}+\delta^{j}_{c}\partial_{a}e^{c}_{i}
-\frac{1}{2}\delta^{j}_{i}\delta^{l}_{c}\partial_{a}e^{c}_{l})\delta^{a}_{j}\nonumber\\
&=&\partial_{a}e^{a}_{i},
\end{eqnarray}
where we have used $\Gamma^{a}_{ac}=\frac{1}{2}\partial_{a}h=\frac{1}{2}\delta^{i}_{c}\partial_{a}e^{c}_{i}$.

In terms of the  linearized spin connection  we define the $U(1)$ connections $\tilde{A}^{j}_{a}$ (one for each internal index)
\begin{eqnarray}\label{conexion}
\tilde{A}^{j}_{a}=\Gamma^{j}_{a}+\beta k^{j}_{a},
\end{eqnarray}
where $\beta$ is an arbitrary constant that corresponds to the Immirzi parameter of the "full" theory \cite{barbero_a,gaul,perez}.
The linearized variables $\tilde{A}^{j}_{a}$ and $e^{a}_{i}$ are canonical, as we shall show. First, observe that the term $e^{a}_{i}\dot{k}^{i}_{a}$ of
 the linearized action becomes
\begin{eqnarray}
e^{a}_{i}\dot{k}^{i}_{a}&=&\frac{1}{\beta}e^{a}_{i}(\dot{\tilde{A}}^{i}_{a}-\dot{\Gamma}^{i}_{a})=
\tilde{e}^{a}_{i}\dot{\tilde{A}}^{i}_{a}- \tilde{e}^{a}_{i}\dot{\Gamma}^{i}_{a},
\end{eqnarray}
with $\tilde{e}^{a}_{i}=\beta^{-1}e^{a}_{i}$. Now, from  the "exact" theory we know that
\cite{thiemann}
\begin{eqnarray*}
E^{a}_{i}\dot{\Gamma}^{i}_{a}=-\frac{1}{2}\varepsilon^{abc}\partial_{a}(\dot{\mathfrak{e}}^{j}_{b}\mathfrak{e}^{i}_{c}\eta_{ij}),
\end{eqnarray*}
whose linearized version, using (\ref{tdl}) (\ref{tnil}) and (\ref{til}), is given by
\begin{eqnarray}
\tilde{e}^{a}_{i}\dot{\Gamma}^{i}_{a}=-\frac{\beta}{2}\varepsilon^{abc}\partial_{a}(\dot{e}^{j}_{b}e^{i}_{c}\eta_{ij})-\beta\delta^{a}_{i}\dot{\Gamma}^{i}_{a}.
\end{eqnarray}
Hence,  $e^{a}_{i}\dot{\Gamma}^{i}_{a}$ can be written down as the sum of total derivative terms which do not  contribute to the equations of motion.
 Therefore, under appropriate boundary conditions, we can make the substitution
\begin{eqnarray}
e^{a}_{i}\dot{k}^{i}_{a}=\tilde{e}^{a}_{i}\dot{\tilde{A}}^{i}_{a}
\end{eqnarray}
in the action. On the other hand, it can be shown that the Poisson algebra between $\tilde{e}^{a}_{i}$ and $\tilde{A}^{i}_{a}$ is given by
\begin{eqnarray}
\{\tilde{e}^{a}_{i}(x),\tilde{A}^{j}_{b}(y)\}&=&\delta^{a}_{b}\delta^{j}_{i}\delta^{3}(x-y)\nonumber\\
\{\tilde{e}^{a}_{i}(x),\tilde{e}^{b}_{j}(y)\}&=&\{\tilde{A}^{i}_{a}(x),\tilde{A}^{j}_{b}(y)\}=0,
\end{eqnarray}
whereby  $\tilde{e}^{a}_{i}$ and $\tilde{A}^{i}_{a}$ form a pair of canonical variables. Hence, the passage to these new variables  constitutes a canonical transformation.

In what follows we shall rewrite all the relevant quantities  of the previous section in terms of the new canonical variables.
Using (\ref{lgav}) the scalar constraint can be written as
\begin{eqnarray}
\mathcal{S}=-2\delta^{k}_{b}\partial^{b}\partial_{c}e^{c}_{k}=2\delta^{k}_{b}\delta^{c}_{m}\varepsilon_{kl}\ ^{m}\partial^{b}\Gamma^{l}_{c}.
\end{eqnarray}
Substituting equation (\ref{conexion}) in the last expression we have
\begin{eqnarray}
 2\delta^{k}_{b}\delta^{c}_{m}\varepsilon_{kl}\ ^{m}\partial^{b}\Gamma^{l}_{c}
=2\delta^{k}_{b}\delta^{c}_{m}\varepsilon_{kl}\ ^{m}\partial^{b}(\tilde{A}^{l}_{c}-\beta k^{l}_{c})=
2\delta^{k}_{b}\delta^{c}_{m}\varepsilon_{kl}\ ^{m}\partial^{b}\tilde{A}^{l}_{c},
\end{eqnarray}
where we have neglected terms proportional to the Gauss constraint. Then, by defining
\begin{eqnarray}
f^{l}_{bc}=\partial_{b}\tilde{A}^{l}_{c}-\partial_{c}\tilde{A}^{l}_{b}
\end{eqnarray}
the scalar constraint can be written as
\begin{eqnarray}
\mathcal{S}=\delta^{k}_{b}\delta^{c}_{m}\varepsilon_{kl}\ ^{m} f^{l}_{bc}\approx0.
\end{eqnarray}

In turn, the vectorial constraint
\begin{eqnarray}
V_{a}=\delta^{c}_{i}\partial_{c}k^{i}_{a}-\delta^{c}_{i}\partial_{a}k^{i}_{c}\approx0
\end{eqnarray}
can be written as
\begin{eqnarray}
V_{a}=\beta^{-1}\delta^{c}_{i}(\partial_{c}\tilde{A}^{i}_{a}-\partial_{a}\tilde{A}^{i}_{c})
- \beta^{-1}\delta^{c}_{i}(\partial_{c}\Gamma^{i}_{a}-\partial_{a}\Gamma^{i}_{c})\approx0,
\end{eqnarray}
where equation (\ref{conexion}) has been used. The second term in this equation vanishes. In fact, from (\ref{celi}), and up to first order in $e^{a}_{i}$ we have
\begin{eqnarray}
\Gamma^{i}_{a}&=&\frac{1}{2}\varepsilon^{ijk}\delta^{b}_{k}
[-\delta^{l}_{a}\delta^{j}_{c}\partial_{b}e^{c}_{l}+\frac{1}{2}\delta^{j}_{a}\delta^{l}_{c}\partial_{b}e^{c}_{l}
+\delta^{l}_{b}\delta^{j}_{c}\partial_{a}e^{c}_{l}-\frac{1}{2}\delta^{j}_{b}\delta^{l}_{c}\partial_{a}e^{c}_{l}
+
\frac{1}{2}\delta^{j}_{a}\delta^{l}_{c}\partial_{b}e^{c}_{l}-\delta^{c}_{j}\delta^{l}_{c}\partial_{b}e^{a}_{l}].
\end{eqnarray}
Hence,
\begin{eqnarray}
\delta^{d}_{i}(\partial_{d}\Gamma^{i}_{a}-\partial_{a}\Gamma^{i}_{d})
&=&
\frac{1}{2}\varepsilon^{dcb}\delta^{l}_{b}\partial_{d}\partial_{a}e^{c}_{l}
+\frac{1}{2}\varepsilon^{dcb}\delta^{l}_{d}\partial_{a}\partial_{b}e^{c}_{l}
-\frac{1}{2}\varepsilon^{dcb}\delta^{l}_{b}\partial_{d}\partial_{a}e^{c}_{l}
+\frac{1}{2}\varepsilon^{cdb}\delta^{l}_{d}\partial_{b}\partial_{a}e^{c}_{l}=0.
\end{eqnarray}

In view of this, the vectorial constraint can be finally cast in the form
\begin{eqnarray}
V_{a}=\tilde{\delta}^{c}_{i}f^{i}_{ca}\approx0,
\end{eqnarray}
with $\tilde{\delta}^{a}_{k}=\beta^{-1}\delta^{a}_{k}$.
Regarding the Gauss constraint, it can be written as
\begin{eqnarray}
G_{i}=\varepsilon_{ij}\ ^{k}k^{j}_{a}\delta^{a}_{k}=\varepsilon_{ij}\ ^{k}\beta k^{j}_{a}\tilde{\delta}^{a}_{k}\approx0.
\end{eqnarray}
But  from equation (\ref{lgav}) we have
\begin{eqnarray}
\partial_{a}\tilde{e}^{a}_{i}+\varepsilon_{ijk}\tilde{\Gamma}^{j}_{a}\tilde{\delta}^{a}_{k}=
\partial_{a}\tilde{e}^{a}_{i}+\varepsilon_{ij}\ ^{k}\Gamma^{j}_{a}\tilde{\delta}^{a}_{k}=0,
\end{eqnarray}
 since the spin connection is invariant under a re-scaling of the triad  \cite{thiemann}.
Introducing the above result in the  expression for the  Gauss constraint we arrive to
\begin{eqnarray}
G_{i}=\partial_{a}\tilde{e}^{a}_{i}+\varepsilon_{ij}\ ^{k}\Gamma^{j}_{a}\tilde{\delta}^{a}_{k}+\varepsilon_{ij}\ ^{k}\beta k^{j}_{a}\tilde{\delta}^{a}_{k}=\partial_{a}\tilde{e}^{a}_{i}+\varepsilon_{ij}\ ^{k}\tilde{A}^{j}_{a}\tilde{\delta}^{a}_{k}\approx0.
\end{eqnarray}

Our final step will be to write $\mathfrak{T}$ as a function of the new canonical variables. From equation
(\ref{ngr}) we can write
\begin{eqnarray}
N\sqrt{q}\ ^{3}R&=&\mathfrak{T}-\nu(\partial^{a}\partial^{b}h_{ab}-\partial^{a}\partial_{b}h),
\end{eqnarray}
which, up to second order, yields
\begin{eqnarray}\label{igual}
(1+\frac{1}{2}h)^{3}R&=&\mathfrak{T}.
\end{eqnarray}
Here, $^{3}R$ is the second order  Ricci scalar, which must be written in terms of the new set of variables.
To this end, we recall the following exact expressions. The curvature tensor is a function of the Christoffel symbols
\begin{eqnarray}\label{3R}
^{3}R^{a}_{bcd}=\partial_{b}^{3}\Gamma^{a}_{cd}-\partial_{c}^{3}\Gamma^{a}_{bd}+\ ^{3}\Gamma^{e}_{cd}\ ^{3}\Gamma^{a}_{be}-\
^{3}\Gamma^{e}_{bd}\ ^{3}\Gamma^{a}_{ce},
\end{eqnarray}
and from equation (\ref{dcce}) we have
\begin{eqnarray}
^{3}\Gamma^{a}_{cd}&=&\mathfrak{e}^{a}_{i}\partial_{c}\mathfrak{e}^{i}_{d} +\mathfrak{e}^{a}_{i}\mathfrak{e}^{j}_{d}\varepsilon_{j}\ ^{i}\ _{k}\Gamma^{k}_{c}.
\end{eqnarray}
Substituting this  in  equation (\ref{3R}) we obtain
\begin{eqnarray}\label{curvaturar}
^{3}R^{a}_{bcd}=\mathfrak{e}^{a}_{i}\mathfrak{e}^{j}_{d}\varepsilon^{i}\ _{lj}\mathcal{F}^{l}_{bc},
\end{eqnarray}
with
\begin{eqnarray}
\mathcal{F}^{l}_{bc}= \partial_{b}\Gamma^{l}_{c}
-\partial_{c}\Gamma^{l}_{b}
+\varepsilon_{k}\ ^{l}\ _{k'}\Gamma^{k}_{c}\Gamma^{k'}_{b}.
\end{eqnarray}

Now we  take  $\Gamma^{i}_{a}$ as  the linearized  connection and replace it according to the canonical transformation that defines the  $U(1)$ connection ${A}^{i}_{a}$
\begin{eqnarray*}
\Gamma^{i}_{a}=\tilde{A}^{i}_{a}-\tilde{k}^{i}_{a},
\end{eqnarray*}
with $\tilde{k}^{i}_{a}=\beta k^{i}_{a}$. This yields, up to the second order in the linear canonical variables
\begin{eqnarray}\label{curvaturaf}
\mathcal{F}^{l}_{bc}&=&F^{l}_{bc}+\varepsilon_{k}\ ^{l}\ _{k'}\tilde{k}^{k}_{c}\tilde{k}^{k'}_{b}+D_{[c}\tilde{k}^{l}_{b]},
\end{eqnarray}
with
\begin{eqnarray}
F^{l}_{bc}=\partial_{b}\tilde{A}^{l}_{c}-\partial_{c}\tilde{A}^{l}_{b}+\varepsilon_{k}\ ^{l}\ _{k'}\tilde{A}^{k}_{c}\tilde{A}^{k'}_{b},
\end{eqnarray}
and
\begin{eqnarray}
D_{c}\tilde{k}^{l}_{b}=\partial_{c}\tilde{k}^{l}_{b}+\varepsilon^{i}\ _{jk}\Gamma^{j}_{c}\tilde{k}^{k}_{b}.
\end{eqnarray}

The next step consist in using equations (\ref{tnil}), (\ref{til}) and (\ref{curvaturaf})  into (\ref{curvaturar}) to build  $^{3}R$ up to the desired order. We have
\begin{eqnarray}\label{escalarr}
^{3}R&=&\delta^{ac}q^{bd}\ ^{3}R^{a}_{bcd}=\mathfrak{e}^{a}_{i}\mathfrak{e}^{j}_{d}\delta^{ac}q^{bd}\varepsilon^{i}\ _{lj}\mathcal{F}^{l}_{bc}
=\varepsilon^{i}\ _{lj}(\delta^{a}_{i}\delta^{b}_{j}+\delta^{a}_{i}e^{b}_{j}+
e^{a}_{i}\delta^{b}_{j})F^{l}_{ba}+\varepsilon^{i}\ _{lj}\varepsilon_{k}\ ^{l}\ _{k'}\delta^{a}_{i}\delta^{b}_{j}\tilde{k}^{k}_{a}\tilde{k}^{k'}_{b}\nonumber\\
&=&2\varepsilon^{i}\ _{lj}\delta^{a}_{i}e^{b}_{j}f^{l}_{ba}+\varepsilon^{i}\ _{lj}\varepsilon_{k}\ ^{l}\ _{k'}\delta^{a}_{i}\delta^{b}_{j}\tilde{A}^{k}_{a}\tilde{A}^{k'}_{b}
+\varepsilon^{i}\ _{lj}\varepsilon_{k}\ ^{l}\ _{k'}\delta^{a}_{i}\delta^{b}_{j}\tilde{k}^{k}_{a}\tilde{k}^{k'}_{b},
\end{eqnarray}
where terms proportional to the scalar constraint have been neglected, and we have defined
$f^{l}_{ba}=\partial_{b}\tilde{A}^{l}_{a}-\partial_{a}\tilde{A}^{l}_{b}$.

Substituting equation (\ref{escalarr}) in (\ref{igual}) we get
\begin{eqnarray}
\mathfrak{T}&=&(1+\frac{1}{2}h)^{3}R=2\varepsilon^{i}\ _{lj}\delta^{a}_{i}e^{b}_{j}f^{l}_{ba}+
\delta^{a}_{i}\delta^{b}_{j}(\tilde{A}^{i}_{a}\tilde{A}^{j}_{b}-\tilde{A}^{j}_{a}\tilde{A}^{i}_{b})
+\delta^{a}_{i}\delta^{b}_{j}(\tilde{k}^{i}_{a}\tilde{k}^{j}_{b}-\tilde{k}^{j}_{a}\tilde{k}^{i}_{b}).
\end{eqnarray}
Finally, the linearized action of the theory in terms of the linear Ashtekar variables results to be
\begin{eqnarray}\label{fin}
S&=&\int d^{4}x(\tilde{e}^{i}_{a}\dot{\tilde{A}}^{i}_{a}-\mathcal{H}-\nu^{d}V_{d}^{L}-\nu S^{L}-N^{i}G_{i}^{L}),
\end{eqnarray}
where
\begin{eqnarray}
\mathcal{S}&=&\delta^{k}_{b}\delta^{c}_{m}\varepsilon_{kl}\ ^{m} f^{l}_{bc}\approx0\nonumber\\
V_{a}&=&\tilde{\delta}^{c}_{i}f^{i}_{ca}\approx0\nonumber\\
G_{i}&=&\partial_{a}\tilde{e}^{a}_{i}+\varepsilon_{ij}\ ^{k}\tilde{A}^{j}_{a}\tilde{\delta}^{a}_{k}\approx0
\end{eqnarray}
are the constraints and
\begin{eqnarray}
\mathcal{H}=2\varepsilon^{i}\ _{jl}\delta^{a}_{i}e^{b}_{j}f^{l}_{ba}
-\delta^{a}_{i}\delta^{b}_{j}(\tilde{A}^{i}_{a}\tilde{A}^{j}_{b}-\tilde{A}^{j}_{a}\tilde{A}^{i}_{b})
-\frac{(\beta^{2}- \frac{1}{4})}{\beta^{2}}\delta^{a}_{i}\delta^{b}_{j}[(\Gamma^{i}_{a}-\tilde{A}^{i}_{a})(\Gamma^{j}_{b}-\tilde{A}^{j}_{b})
-(\Gamma^{j}_{a}-\tilde{A}^{j}_{a})(\Gamma^{i}_{b}-\tilde{A}^{i}_{b})].
\end{eqnarray}
is the  Hamiltonian density.

\section{Concluding Remarks}

The expressions for the Hamiltonian and the constraints obtained in the present article coincide with those of reference  \cite{dibartolo},
were the procedure for attaining the linearized Ashtekar formulation was different to the one we followed. In our case, we first made the linearization
from the first order ADM action and then performed the passage to "linear" Ashtekar variables. In the previous works, instead, the linearization was performed after having
 formulated the "full" theory in the Ashtekar new variables. Nevertheless, there is a subtle conceptual difference between both approaches, regarding the linearization point,
which is worth mentioning. One might wonder about where comes the linear Hamiltonian from, since in the full theory there is no Hamiltonian at all, but just constraints.
The answer is that the Hamiltonian comes from the multiplication of the $0-th$ order lagrange multipliers (the lapsus and the shift functions) times the quadratic part of
the scalar and vectorial constraints.  Now, at the level of the ADM linear action, it is obvious what these $0-th$ orders should be: it suffices to see how the lapse and
the shift relate with the metric components (equations (\ref{tres})) to get the answer (equations (\ref{ventiuno})). We believe that this point can be better
understood within the
approach discussed in this article than in the standard one. In fact, one could conceive different linearizations starting from the "full" Ashtekar formulation, in which
there is no Hamiltonian at all \cite{varadarajan}. This amounts to taking a linear theory different from the Fierz-Pauli one, which could also be consistent, but
that could lead to different physical predictions.

Finally, it is interesting to notice that the linear theory, unlike the "full" one, could admit different versions of the "Loop Representation", in the following sense.
Being an Abelian theory (like the Maxwell theory), there exist the possibility of both an "electric" and a "magnetic" representation. In the former, the linearized triad
 would act as the loop form factor (i.e. the "loop coordinate"), the linear Ashtekar connection taking the role of the "path derivarive" and its curl acting as a "loop
derivative". But since the linear theory is dual (in the "electric-magnetic" sense) (see reference \cite{hennaux}), these roles could be interchanged,
just as in Maxwell theory.
On the other hand, and closely related with the previous discussion, it seems possible to consider the introduction of a  "Loop Representation" at  stages previous to
the introduction of the linearized Ashtekar variables. For instance, the canonical pairs $(e^{a}_{i},k^{i}_{a})$ could serve as a starting point for doing this. These
aspects are currently under work.

\end{document}